\documentclass[aps,prl,twocolumn,groupedaddress,showpacs]{revtex4}

\usepackage[dvips]{graphicx}
\usepackage{epsfig}
\usepackage{amssymb}
\usepackage{amsmath}
\usepackage{color}

\begin{document}


\title{Mean-field energy-level shifts and dielectric properties of strongly polarized Rydberg gases}


\author{V. Zhelyazkova, R. Jirschik and S. D. Hogan}

\affiliation{Department of Physics and Astronomy, University College London, Gower Street, London WC1E 6BT, U.K.}


\date{\today}

\begin{abstract} 
Mean-field energy-level shifts arising as a result of strong electrostatic dipole interactions within dilute gases of polarized helium Rydberg atoms have been probed by microwave spectroscopy. The Rydberg states studied had principal quantum numbers $n=70$ and~72, and electric dipole moments of up to 14\,050~D, and were prepared in pulsed supersonic beams at particle number densities on the order of $10^{8}$~cm$^{-3}$. Comparisons of the experimental data with the results of Monte Carlo calculations highlight effects of the distribution of nearest-neighbor spacings in the pulsed supersonic beams, and the dielectric properties of the strongly polarized Rydberg gases, on the microwave spectra. These observations reflect the emergence of macroscopic electrical properties of the atomic samples when strongly polarized.

\end{abstract}

\pacs{32.80.Rm}

\maketitle

Rydberg states of atoms and molecules can possess large static electric dipole moments. These dipole moments exceed $10\,000$~D for states with principal quantum numbers $n \gtrsim 50$, and scale with $n^2$. They give rise to strong linear Stark shifts in external electric fields~\cite{gallagher94a}, and allow for control of the translational motion of a wide range of neutral atoms and molecules~\cite{hogan16a}, composed of matter~\cite{yamakita04a,vliegen04a,hogan08a,hogan09a} and antimatter~\cite{deller16a}, using inhomogeneous electric fields. When prepared in states with these large dipole moments, gases of Rydberg atoms can also exhibit strong electrostatic dipole-dipole interactions~\cite{gallagher08a,zhelyazkova15a,wang15a}. These inter-particle interactions cause energy shifts of the Rydberg states, and, as demonstrated here, can modify the dielectric properties of the gases. The observation, and spectroscopic characterization, of the effects of these electrostatic interactions opens opportunities for using beams of Rydberg atoms or molecules as model systems with which to study many-body processes~\cite{carr09a}, including, e.g., resonant energy transfer~\cite{smith78a}, or surface ionization~\cite{hill00a,lloyd05a,gibbard15a}. It is also of importance in the elucidation of effects of dipole interactions in Rydberg-Stark deceleration and trapping experiments~\cite{seiler16a}, in experiments involving long-range Rydberg molecules~\cite{green00a,bendkowsky09a} possessing large static electric dipole moments~\cite{booth15a}, and in applications of Rydberg states as microscopic antennas for the detection of low-frequency electric fields~\cite{zhelyazkova15a,miller16a}.

Microwave spectroscopy of the effects of van der Waals interactions~\cite{raimond81a,park11a,teixeira15a}, and resonant dipole-dipole interactions~\cite{afrousheh04a,park16a} in cold Rydberg gases have been reported previously and provided important insights into these interacting few-particle systems. The electrostatic interactions that occur between Rydberg atoms in the outermost Stark states with the largest electric dipole moments, $\mu_{\mathrm{max}}\simeq(3/2)n^2 e a_0$, represent the extreme case of the dipole-dipole interaction for each value of $n$. The interaction potential, $V_{\mathrm{dd}}$, between a pair of atoms with electric dipole moments, $\mu_{1}=|\vec{\mu}_{1}|$ and $\mu_{2}=|\vec{\mu}_{2}|$, aligned with an electric field, $\vec{F}$, can be expressed as~\cite{gallagher08a}
\begin{eqnarray}
V_{\mathrm{dd}} &=& \frac{\mu_{1}\mu_{\mathrm{2}}}{4\pi\epsilon_0 R^3}\left(1- 3\cos^2\theta \right),
\end{eqnarray}
where $R=|\vec{R}|$ is the inter-atomic distance, $\theta$ is the angle between $\vec{R}$ and $\vec{F}$, and $\epsilon_0$ is the vacuum permittivity. For an isotropic, aligned ensemble of dipoles the average interaction energy is non zero, and leads to mean-field energy-level shifts within the ensemble. 

The microwave spectroscopic studies reported here represent a direct measurement of these mean-field shifts in gases of helium (He) excited to Rydberg states with $n=70$ and electric dipole moments up to $\mu_{1} = \mu_{70} = 12\,250$~D. In the highest density regions of these gases, the dielectric properties differ from those in free space and give rise to changes in the microwave spectra which reflect the emergence of macroscopic electrical properties of the medium. This work is complimentary to recent studies of effects of electrostatic dipole interactions on particle motion in laser cooled samples of rubidium~\cite{thaicharoen16a}, and observations of optical bistability arising from mean-field shifts in gases of Rydberg atoms interacting via resonant dipole interactions~\cite{carr13a}.

\begin{figure}
\begin{center}
\includegraphics[width = 0.46\textwidth, angle = 0, clip=]{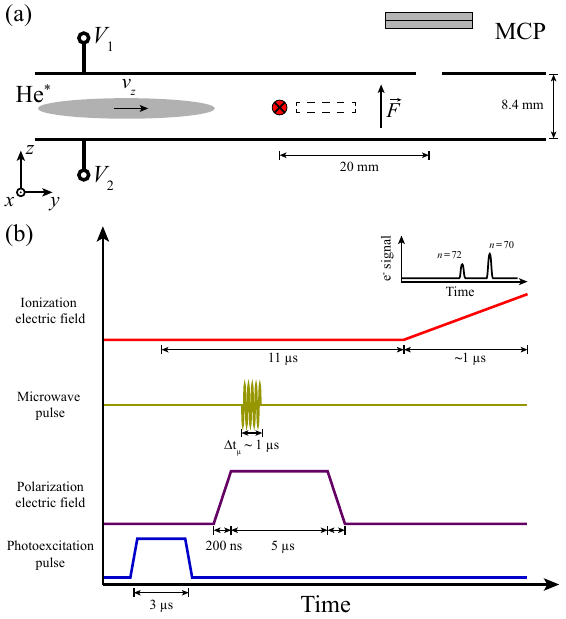}
\caption{(a) Schematic diagram of the photoexcitation and detection region of the experimental apparatus. The position of laser photoexcitation and the approximate spatial extent of the Rydberg atom ensemble in the $y$ dimension during microwave interrogation are indicated by the red circle and dashed rectangle, respectively. (b) Sequence of events in each experimental cycle, including laser photoexcitation, Rydberg state polarization, microwave interrogation, and state-selective electric field ionization (not to scale).}
\label{fig1}
\end{center}
\end{figure}

The experiments were performed in pulsed supersonic beams of metastable He with a mean longitudinal speed of $\sim2000$~m\,s$^{-1}$~\cite{zhelyazkova15b}. The atoms were prepared in the metastable 1s2s\,$^3$S$_1$ level in an electric discharge at the exit of a pulsed valve operated at a repetition rate of 50~Hz~\cite{halfman00a}. After collimation by a 2~mm diameter skimmer and the deflection of stray ions produced in the discharge, the beam entered between a pair of parallel 70~mm $\times$ 70~mm copper electrodes, separated in the $z$ dimension by 8.4~mm, as indicated in Figure~\ref{fig1}(a). At the mid-point between the two electrodes the atomic beam was crossed at right-angles by two co-propagating, frequency stabilized, cw laser beams at wavelengths of $\lambda_{\mathrm{UV}}=388.975$~nm and $\lambda_{\mathrm{IR}}= 785.946$~nm to excite the metastable atoms to Rydberg states by the 1s2s\,$^3$S$_1\rightarrow$1s3p\,$^3$P$_2\rightarrow$1s70s\,$^3$S$_1$ two-photon excitation scheme. The laser beams were focussed to full-width-at-half-maximum (FWHM) beam waists of $\sim100~\mu$m. By applying a pulsed potential, $V_1$, to the upper electrode in Figure~\ref{fig1}(a) to bring the atoms in the photoexcitation region into resonance with the lasers for $3~\mu$s [see Figure~\ref{fig1}(b)], 6~mm-long ensembles of Rydberg atoms were generated in each cycle of the experiment. 

\begin{figure}
\begin{center}
\includegraphics[width = 0.48\textwidth, angle = 0, clip=]{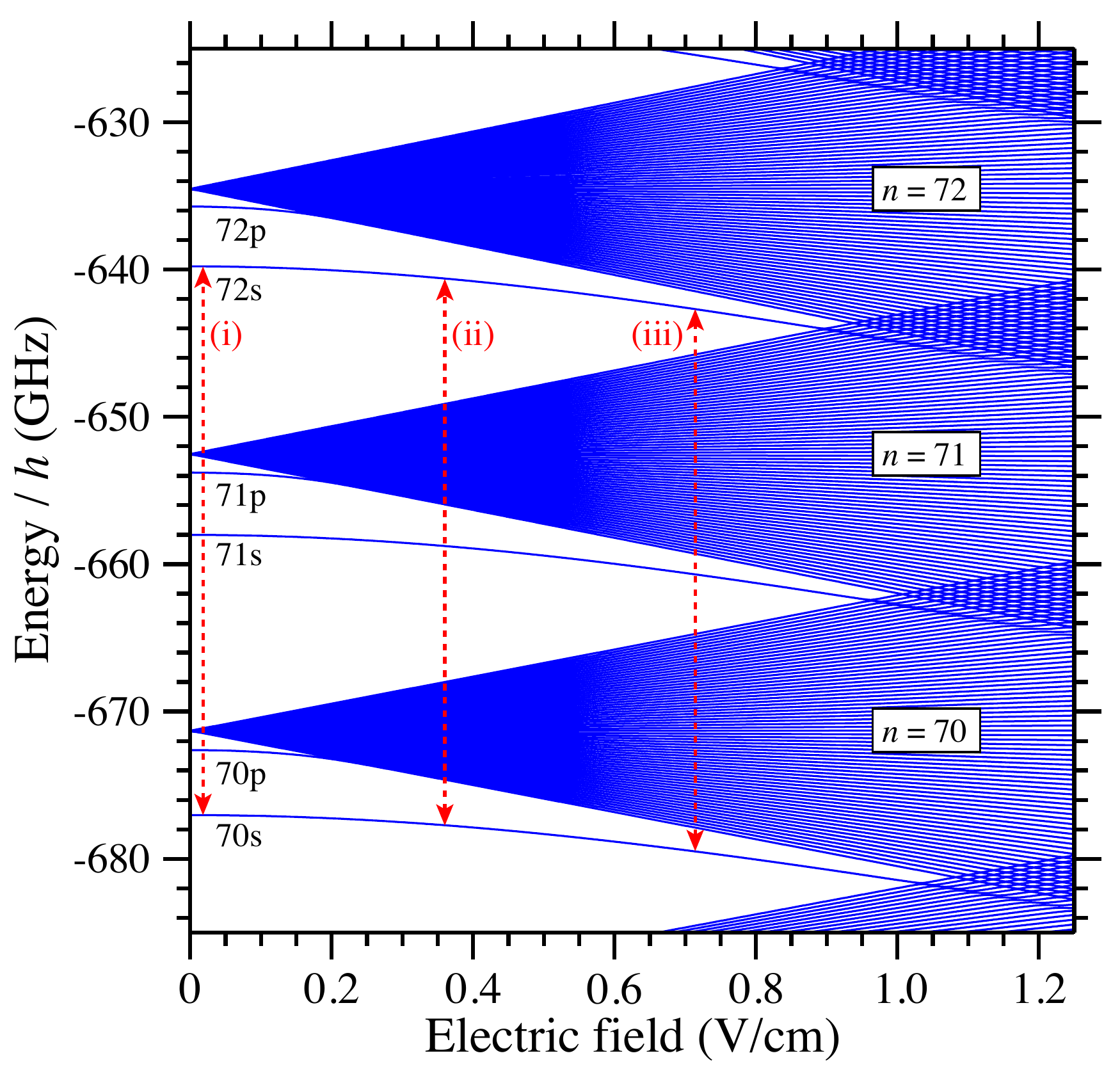}
\caption{Stark structure of the triplet $m_{\ell}=0$ Rydberg states of He with values of $n$ ranging from 70 to 72. Experiments were performed following laser photoexcitation of the 70s state in zero electric field and the subsequent polarization of the excited ensemble of atoms in fields of (i) 18~mV\,cm$^{-1}$, (ii) 358~mV\,cm$^{-1}$, and (iii) 714~mV\,cm$^{-1}$ as indicated.}
\label{fig2}
\end{center}
\end{figure}

After photoexcitation, the electric field was switched again to polarize the 1s70s\,$^3$S$_1$ atoms [see Figure~\ref{fig1}(b)] before interactions within the excited ensemble were probed by microwave spectroscopy on the single photon transition at $\sim37$~GHz between the states that evolve adiabatically to the 1s70s\,$^3$S$_1$ and 1s72s\,$^3$S$_1$ levels in zero electric field. These states have quantum defects of $\delta_{70\mathrm{s}}\simeq\delta_{72\mathrm{s}}\simeq0.296\,664$~\cite{drake99a} and exhibit quadratic Stark energy shifts. Because the laser photoexcitation process was separated in time from microwave interrogation, the Rydberg atom number density, $n_{\mathrm{Ry}}$, could be adjusted by controlling the IR laser intensity, without affecting any other experimental parameters. After traveling 20~mm from the position of laser photoexcitation, the Rydberg atoms were detected by state-selective ramped-electric-field ionization beneath a 3~mm-diameter aperture in the upper electrode [see Figure~\ref{fig1}(a)] by applying a slowly-rising ionization potential, $V_2$, to the lower electrode. The resulting electrons were detected on a microchannel plate (MCP) detector located above this electrode. The rise time of the pulsed ionization electric field was selected to separate in time the electron signal from the $n=70$ and $n=72$ states. 

\begin{figure*}
\begin{center}
\includegraphics[width = 0.75\textwidth, angle = 0, clip=]{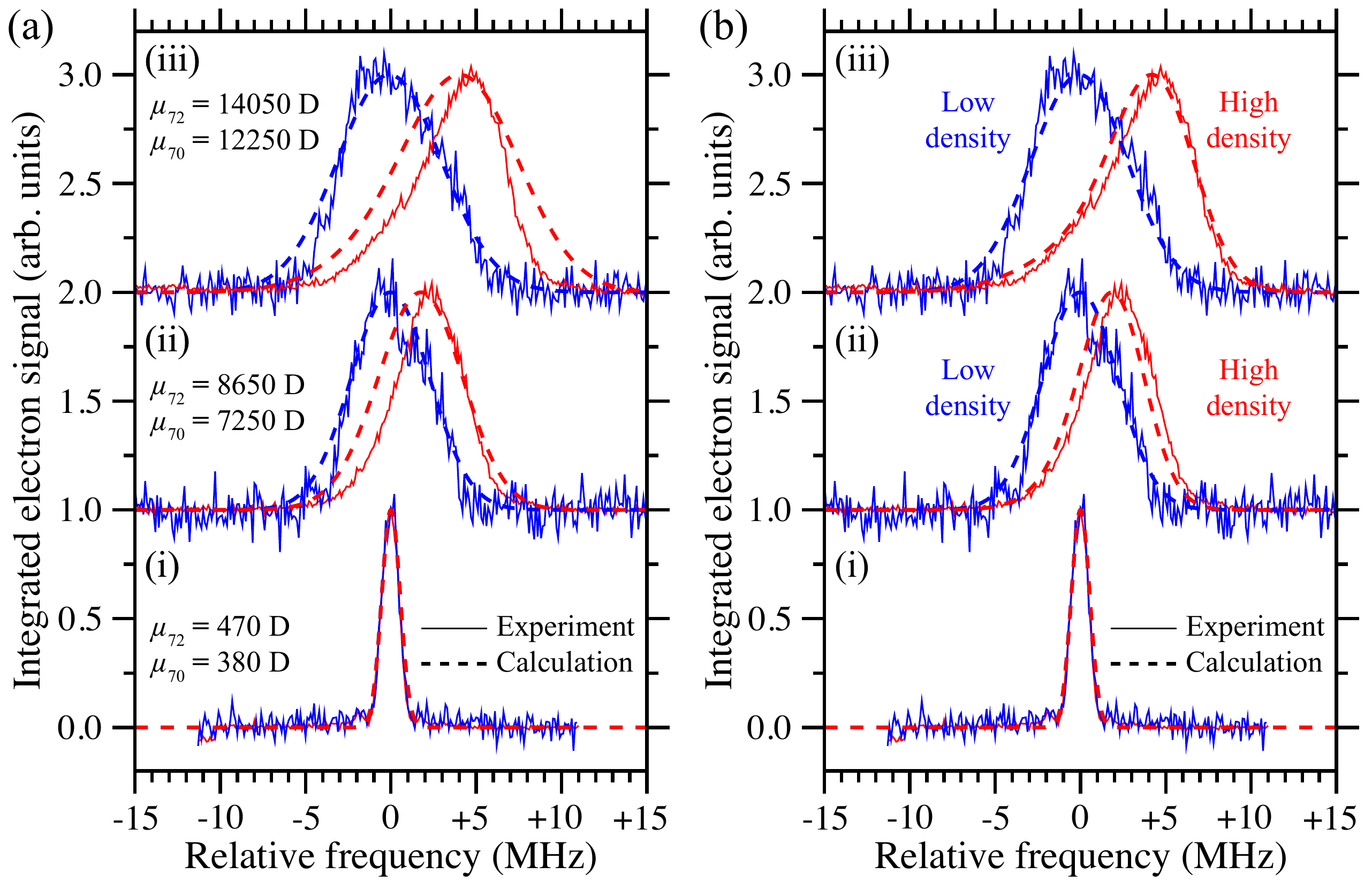}
\caption{Experimentally recorded (continuous curves) and calculated (dashed curves) spectra of transitions between the Stark states that evolve adiabatically to the 1s70s\,$^3$S$_1$ and 1s72s\,$^3$S$_1$ levels in zero electric field. The experiments were performed at low (blue curves) and high (red curves) Rydberg atom number densities, and spectra were recorded in polarizing electric fields of (i)~$18$~mV\,cm$^{-1}$, (ii) 358~mV\,cm$^{-1}$, and (iii) 714~mV\,cm$^{-1}$, for which $\mu_{70} = 380$~D, 7250~D, and 12250~D ($\mu_{72} = 470$~D, 8650~D, and 14050~D), respectively. The calculated spectra are presented (a) without, and (b) with effects of the local polarization of the medium on the macroscopic dielectric properties accounted for. The relative microwave frequencies on the horizontal axes are displayed with respect to the transition frequencies [(i) 37.2413~GHz, (ii) 37.1068~GHz, and (iii) 36.8094~GHz] recorded at low Rydberg atom number density.}
\label{fig3}
\end{center}
\end{figure*}

The mean-field energy-level shifts of the polarized Rydberg gases were observed directly by microwave spectroscopy. To achieve this, reference measurements were first made in which the polarizing electric field was switched after photoexcitation from $0\pm1$~mV\,cm$^{-1}$ to 18~mV\,cm$^{-1}$, inducing electric dipole moments of $\mu_{70}=380$~D in the 1s70s\,$^3$S$_1$ atoms. Here, and in the following, the dipole moments referred to represent the absolute value of the first derivative of the Stark energy with respect to the electric field. The $n=70\rightarrow n=72$ transition ($\mu_{72}=470$~D), indicated by the dashed line labelled~(i) in Figure~\ref{fig2}, was then driven by a 1~$\mu$s-long microwave pulse [see Figure~\ref{fig1}(b)]. The corresponding Fourier-transform-limited spectrum of the integrated electron signal associated with the $n=72$ state is displayed in Figure~\ref{fig3}(a-i) (continuous curves). Measurements were performed at two Rydberg atom densities. The low (high) density data were recorded following the excitation of $N_{\mathrm{Ry}}$ ($16\,N_{\mathrm{Ry}}$) Rydberg atoms, and are displayed in blue (red) in the figure. For the small electric dipole moments induced in this case, no density-dependent energy-level shifts were observed.

To enhance the effects of electrostatic dipole interactions within the Rydberg gases, subsequent spectra were recorded with further polarization. Increasing the electric field to 358~mV\,cm$^{-1}$ (714~mV\,cm$^{-1}$) resulted in induced electric dipole moments of $\mu_{70}=7250$~D ($\mu_{70}=12250$~D), and $\mu_{72}=8650$~D ($\mu_{72}=14050$~D). In the spectra recorded in these fields, and displayed in Figure~\ref{fig3}(a-ii) and (a-iii), respectively, a significant dependence on the Rydberg atom density was observed. Comparison of the spectra in Figure~\ref{fig3}(a-ii) and (a-iii) with those in Figure~\ref{fig3}(a-i) reveals four notable features. When the atoms are more strongly polarized the spectral profiles: (1) become significantly broader than those in Figure~\ref{fig3}(a-i), even at low density; (2) exhibit density-dependent frequency shifts; (3) become asymmetric at high density with a sharp cut-off in intensity at higher microwave transition frequencies; and (4) display signatures of spectral narrowing at high density, in particular in Figure~\ref{fig3}(a-iii).

To aid in the interpretation of the spectra, Monte Carlo calculations were performed in which the electrostatic dipole-dipole interactions within the many-particle system were treated. In these calculations, ensembles of Rydberg atoms were generated with randomly assigned positions. These ensembles had Gaussian spatial distributions with a FWHM of 100~$\mu$m (50~$\mu$m) in the $x$ ($z$) dimension, and flat top distributions with a length of 3~mm in the $y$ dimension. These distributions represent the spatial intensity profile of the laser beams in the $z$ dimension, the spatial distribution of atoms with Doppler shifts within the spectral width ($\Delta\nu_{\mathrm{FWHM}}\simeq15$~MHz) of the 1s2s\,$^3$S$_1\rightarrow$1s3p\,$^3$P$_2$ transition in the $x$ dimension, and half of the length of the ensemble of atoms excited in the $y$ dimension.  

For each atom in the calculation the difference between the sum of the dipole-dipole interaction energies with all other atoms when it was (1) in the $n=70$ state, i.e., $\mu_1 = \mu_2 = \mu_{70}$, and (2) in the $n=72$ state, i.e., $\mu_1 = \mu_{72}$, and $\mu_2 = \mu_{70}$, was determined. To account for the sharp cut-off in the spectral profiles at high transition frequencies in the data recorded at high density in Figure~\ref{fig3}(a-ii) and (a-iii), it was necessary to impose a lower limit, $R_{\mathrm{min}}$, on the nearest-neighbor spacing between pairs of Rydberg atoms in the calculations. This minimum inter-atomic spacing is a consequence of the expansion of the pulsed supersonic beams as they propagate from the valve to the photoexcitation region in the experiments, a distance of 210~mm, and reflects the collision-free environment characteristic of these beams~\cite{lubman82a,smalley77a}. From a global fit to all of the experimental data in Figure~\ref{fig3} the most appropriate value of $R_{\mathrm{min}}$ was found to be $11.5~\mu$m. We note that because the $C_6$, van der Waals coefficient of the 1s70s\,$^3$S$_1$ level in He is $\sim9.9$~GHz\,$\mu\mathrm{m}^6$, blockade at laser photoexcitation does not play a major role at the laser resolution in the experiments for atoms separated by more then $\sim3.5~\mu$m and therefore does not dominate this value of $R_{\mathrm{min}}$. Under these conditions, the distribution of $n=70\rightarrow n=72$ transition frequencies was determined for each set of electric dipole moments (i.e., for each electric field strength) for atoms located within $\pm0.75$~mm of the center of the ensemble in the $y$-dimension to avoid edge effects. The resulting data were convoluted with a Gaussian spectral function with a FWHM of 1~MHz corresponding approximately to the Fourier transform of the microwave pulses. 

When the polarization of the individual atoms is increased they become more sensitive to low-frequency electrical noise~\cite{zhelyazkova15a}. This sensitivity led to the broadening of the resonances in the low-density regime in Figure~\ref{fig3}(a-ii) and (a-iii). The effect of this spectral broadening was introduced in the calculations as a perturbing electric field which caused a relative shift in the energy of the $n=70$ and $n=72$ Stark states, proportional to their electric dipole moments. The spectra calculated following the generation of low density samples containing $N_{\mathrm{Ry}}=2\,813$ atoms and including the contributions from this electric field noise are displayed in Figure~\ref{fig3}(a) (blue dashed curves). Comparing the experimental data with the results of these calculations indicates that the broadening observed was commensurate with white noise with a root-mean-square amplitude of $F_{\mathrm{noise}} = 2$~mV\,cm$^{-1}$.

In the more polarized gases, increasing the Rydberg atom density is seen to shift the resonant microwave transition frequencies by $\sim+3$~MHz and $\sim+5$~MHz in Figure~\ref{fig3}(a-ii) and (a-iii), respectively, from those recorded at low density. These density-dependent changes indicate that atom-atom interactions dominate the spectral broadening caused by $F_{\mathrm{noise}}$. The mean-field shifts in transition frequency were found to agree with those seen in the results of the calculations upon increasing the number of excited atoms by a factor of~16, matching the experiments, to $N_{\mathrm{Ry}}=45\,000$ [red dashed curves in Figure~\ref{fig3}(a)]. The positive shifts in transition frequency with increasing density are a consequence of the predominantly repulsive electrostatic dipole interactions within the elongated ensemble of Rydberg atoms in the experiments, and the larger energy shift of the more polar, $n=72$ state under these conditions.

\begin{figure}
\begin{center}
\includegraphics[width = 0.4\textwidth, angle = 0, clip=]{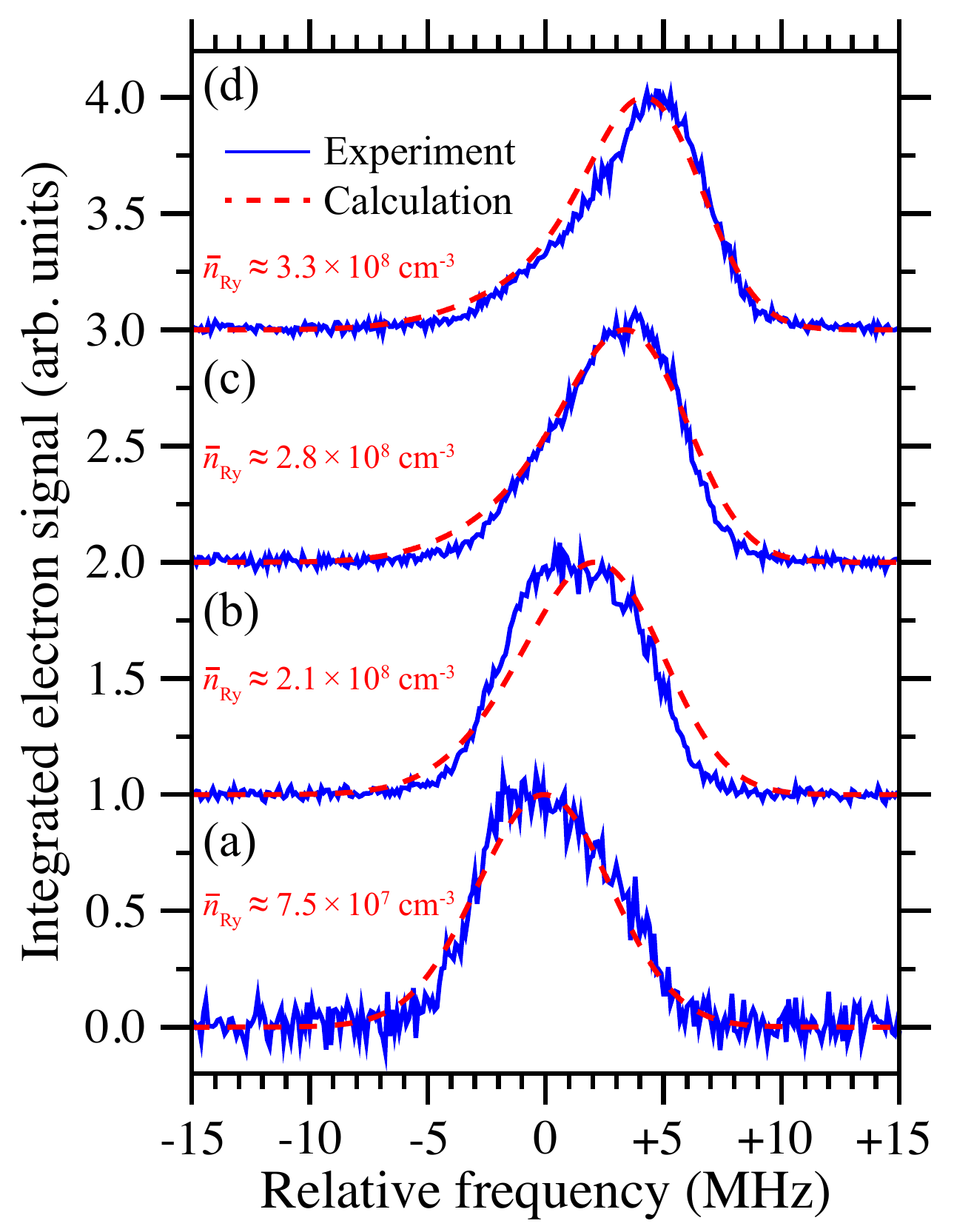}
\caption{Experimentally recorded (continuous curves) and calculated (dashed curves) dependence of the $n=70\rightarrow n=72$ spectra of the most strongly polarized ensemble of atoms, for which $\mu_{70}=12\,250$~D and $\mu_{72}=14050$~D, on the Rydberg atom density.}
\label{fig4}
\end{center}
\end{figure}

However, the spectral narrowing observed at high density in the experimental data is not seen in the calculations in Figure~\ref{fig3}(a). To account for this it was necessary to consider the contribution of the local polarization of the Rydberg gases on their dielectric properties~\cite{panofsky62a}. In a simple model of the dielectric Rydberg gas, the local polarization $P_{\mathrm{loc}} = n_{\mathrm{loc}}\,\mu_{70}$ ($n_{\mathrm{loc}}$ is the Rydberg atom number density within a sphere of radius 25~$\mu$m surrounding each atom) can be considered to shield each atom from the laboratory electrical noise, reducing it to $F_{\mathrm{loc}}=\mathrm{max}(\{0, F_{\mathrm{noise}}-S_{\kappa}P_{\mathrm{loc}}/\epsilon_0\})$, where $S_{\kappa}$ is a constant factor that accounts for the geometry of the ensemble of atoms and the value of $R_{\mathrm{min}}$ in the experiments. For $S_{\kappa}=0.1$, there is excellent agreement between the results of the calculations and the experimental data as can be seen in Figure~\ref{fig3}(b). This indicates the emergence of macroscopic electrical properties of the Rydberg gas when strongly polarized. In the experiments, free He$^+$ ions, generated by photoionization of Rydberg atoms, would give rise to spectral broadening, or, if they increase the strength of the local electric fields within the ensembles of atoms, shifts of the spectral features toward lower transition frequencies (see Figure~\ref{fig2}). They are not expected to contribute to spectral narrowing such as that observed in Figure~\ref{fig3}(a,ii-iii) and (b,ii-iii). For these reasons, and because we do not observe free ions or electrons in the process of ionization of the Rydberg atoms, we conclude that they do not contribute significantly to the experimental observations. 

Further comparisons of experimental and calculated spectra of the most polarized atoms [Figure~\ref{fig3}(b,iii)] over a range of Rydberg atom densities are presented in Figure~\ref{fig4}. From these data, it can be seen that the experimental spectra, recorded upon increasing $N_{\mathrm{Ry}}$ by factors of 5, 9.5 and 16 [Figure~\ref{fig4}(b), (c) and (d), respectively] from the initial low density case [Figure~\ref{fig4}(a)], agree well with the calculated spectra containing equivalent proportions of atoms, i.e., $N_{\mathrm{Ry}}=2\,813,\; 14\,063,\; 26\,719$, and $45\,000$. From the results of the calculations the mean Rydberg atom number densities, $\overline{n}_{\mathrm{Ry}}$, within the samples could be determined to range from $\overline{n}_{\mathrm{Ry}} = 7.5\times10^{7}$~cm$^{-3}$ to $3.3\times10^{8}$~cm$^{-3}$, as indicated. In these strongly polarized gases the laboratory electric field noise, $F_{\mathrm{noise}}$, is completely screened when $S_{\kappa}P_{\mathrm{loc}}/\epsilon_0 \geq$ 2~mV\,cm$^{-1}$, i.e., when $n_{\mathrm{loc}} \geq 4.5\times10^{8}$~cm$^{-3}$, and the absolute energy-level shift of each atom is $\sim20$~MHz.

In conclusion, we have carried out spectroscopic studies of mean-field energy-level shifts in strongly polarized Rydberg gases by driving microwave transitions between Rydberg-Stark states with similar electric dipole moments. The resulting spectra yield detailed information on the spatial distributions of the atoms. Spectral narrowing observed at high number density is attributed to changes in the local dielectric properties within the medium from those in free space, and reflects the macroscopic electrical properties of the atomic samples that emerge under these conditions.

\begin{acknowledgments}
This work was supported financially by the Department of Physics and Astronomy and the Faculty of Mathematical and Physical Sciences at University College London, and the Engineering and Physical Sciences Research Council under Grant No.~EP/L019620/1.
\end{acknowledgments}

\end{document}